\documentstyle[12pt,epsf]{article}
\newcommand{\nll}{\nonumber \\}

\newcommand{\bq}{\begin{equation}}
\newcommand{\eq}{\end{equation}}
\newcommand{\ba}{\begin{eqnarray}}
\newcommand{\ea}{\end{eqnarray}}
\newcommand{\req}[1]{(\ref{#1})}
\newcommand{\nobody}{\rule{0ex}{1ex}}

\newcommand{\nobodyfrac}{\frac{\nobody}{\nobody}}
\begin{document}
\voffset -2cm
\begin{flushleft}
LMU-01/97
\end{flushleft}
\begin{center}
\vspace{1.5cm}\hfill\\
{\LARGE 
Remark on $Z'$ limits at hadron colliders
}\vspace{1cm}\\
A. Leike\footnote{Supported by the EC-program CHRX-CT940579}\\
Ludwigs--Maximilians-Universit\"at, Sektion Physik, Theresienstr. 37,\\
D-80333 M\"unchen, Germany\\
E-mail: leike@graviton.hep.physik.uni-muenchen.de
\end{center}
%
\begin{abstract}
\noindent
Simple estimates for $Z'$ exclusion limits and $Z'$ model
measurements in $pp\ (p\bar p)$ collisions are derived.
Due to properties of the structure functions, the $Z'$ exclusion
limits depend only logarithmically on the $Z'$ couplings to fermions
and on the integrated luminosity.
The predicted scaling of $Z'$ exclusion limits and errors of $Z'$ measurements
with the c.m. energy and luminosity allows an easy extrapolation of
existing analyses to other colliders.
 \vspace{1cm}
\end{abstract}
%
%
It is well known that $Z'$ exclusion limits from $e^+e^-$ collisions
vary strongly with the $Z'$ model, while the $Z'$ constraints from
$pp$ and $p\bar p$ collisions show only a weak model dependence
\cite{sammelref} - \cite{zpmi}. 
The mechanisms leading to $Z'$ limits in hadron collisions and in $e^+e^-$
collisions are essentially different.
The $Z'$ is detected through {\it indirect} interference effects between the
$Z'$ contributions and the SM contributions in $e^+e^-$ collisions.
It is detected through {\it direct} production in $pp$ or $p\bar p$
collisions. 
However, this cannot be the origin of the weak model dependence of the $Z'$
exclusion limits from hadron collisions because the cross section of
{\it direct} $Z'$ production depends on the fourth power of the model
dependent $Z'$ couplings to SM fermions, while the {\it indirect} $Z'$ limits
depend only on the square of these couplings.

In this paper, we show that the weak model dependence of the $Z'$
exclusion limits from hadron collisions is due to properties of the structure 
functions, which lead to an effective {\it logarithmic} dependence of the
$Z'$ limits on the $Z'$ couplings.
The decay mode of the $Z'$ does not influence this dependence.
Therefore, we assume for simplicity that the $Z'$ is detected
through a muon pair. 
Our derivation is based on the assumption that the $Z'$ is produced by
a quark antiquark pair. 
The considered reaction is much less sensitive to $ZZ'$ mixing than
$e^+e^-$ collisions at the $Z$ peak.
Therefore, $ZZ'$ mixing effects can be neglected.

We now derive the scaling law of $Z'$ limits with the c.m. energy $\sqrt{s}$
and the integrated luminosity $L$ approximating the Born cross section,
\ba
\label{ppzpborn}
\sigma_T\left(pp(p\bar p)\rightarrow (\nobodyfrac\gamma,Z,Z')X\rightarrow
\mu^+\mu^-X\right)\hspace{8cm}\ ~\\
= \sum_{q} \int_0^1dx_1\int_0^1dx_2
\sigma_T(sx_1x_2;q\bar q\rightarrow \mu^+\mu^-)
G_T^q(x_1,x_2,M^2_{Z'})\theta(x_1x_2s-M_\Sigma^2),\nonumber
\ea
where $G_T^q(x_1,x_2,M_{Z'}^2)$ depends on the structure functions
of the partons $q$ and $\bar q$,
\bq
\label{gdists}
G_T^q(x_1,x_2,M^2_{Z'}) = q(x_1,M_{Z'}^2)\bar q(x_2,M_{Z'}^2)
+\bar q(x_1,M_{Z'}^2) q(x_2,M_{Z'}^2),
\eq
and $M_\Sigma$ is the sum of the masses of the final particles. 
We assume that the $Z'$ signal is well above the
SM background in the appropriate region of the invariant energy of the
muon pairs. 
This condition is fulfilled in the $E_6$ GUT, the left-right theory
and in many other GUT's \cite{e6}.

The resonating $Z'$ propagator of the subprocess $\sigma_T(Q^2;q\bar
q\rightarrow \mu^+\mu^-)$ can then be treated in the narrow width
approximation, 
\bq
\frac{Q^4}{|Q^2-M^2+iM\Gamma|^2}\longrightarrow
\delta(Q^2-M^2)\frac{\pi M^4}{M\Gamma},
\eq
leading to \cite{e6}
\ba
\label{ppzpnwa}
\sigma_T\left(pp(p\bar p)\rightarrow (\nobodyfrac\gamma,Z,Z')X\rightarrow
\mu^+\mu^-X\right)\hspace{8cm}\ ~\nll
= \frac{4\pi^2}{3s}\frac{\Gamma_{Z'}}{M_{Z'}}
Br(Z'\rightarrow \mu^+\mu^-)
\sum_{q} Br(Z'\rightarrow q\bar q)
f^q\left(\frac{\sqrt{s}}{M_{Z'}},M_{Z'}^2 \right),
\ea
\bq
\label{fdef}
\mbox{with}\hspace{1cm}  f^q\left(r_z,M_{Z'}^2\right)
=\int_{1/r_z^2}^1\frac{dx}{x}G_T^q\left(x,\frac{1}{sxr_z^2},M^2_{Z'}\right)
\mbox{\ \ and\ \ } r_z=\frac{\sqrt{s}}{M_{Z'}}.
\eq

A numerical inspection of the function $f^q\left(r_z,M_{Z'}^2\right)$
shows that it has only a very weak dependence on $M_{Z'}^2$ in the region we
are interested in, i.e. 
$f^q\left(r_z,M_{Z'}^2\right)\approx f^q\left(r_z\right)$.   
Furthermore, the functions for different quarks $q=u,d$ differ mainly by a
constant factor. 
For our purposes, we can make the following replacement in equation
\req{ppzpnwa}, 
\bq
\label{fzdef}
\sum_{q} Br(Z'\rightarrow q\bar q)
f^q\left(\frac{\sqrt{s}}{M_{Z'}},M_{Z'}^2 \right)=
f^u\left(\frac{\sqrt{s}}{M_{Z'}}\right)
\left[ Br(Z'\rightarrow u\bar u) + \frac{1}{C_{ud}} Br(Z'\rightarrow d\bar d)
\right].
\eq
Remembering the $pp$ and $p\bar p$ colliders under discussion \cite{godfrey},
we see that the functions $f^q(r_z)$ are needed only in a narrow interval of
$r_z$, i.e. $3<r_z<5$ for $pp$ collisions and $2<r_z<3.5$ for $p\bar
p$ collisions.  
In these regions, we have 
$C_{ud}=f^u(r_z)/f^d(r_z)\approx 2\ (25)$ for $pp\ (p\bar p)$ collisions. 
Therefore, the $Z'$ search has a reduced sensitivity to $Z'd\bar d$ couplings,
especially in $p\bar p$ collisions.

The integral defining the function $f^u(r_z)$ could be approximated by
the function $r_z^a(r_z-1)^b$, which takes into accound the
parametrization of the structure functions.
For our purposes, we would prefer an approximation with a function,
which can be inverted analytically.
It turns out that $f^u(r_z)$ can be fitted by an
exponential function in the relevant interval of $r_z$,
\bq
\label{fapprox2}
f^u(r_z)\approx Ce^{-A/r_z},\ \ \ 
C=600\ (300),\ A=32\ (20)\mbox{\ \ for\ \ }pp\ (p\bar p)\mbox{\ collisions.}
\eq
The approximation \req{fapprox2} and the exact calculation \req{fdef} of
$f^u(r_z)$ are shown in figure~1.
Note that the fit works satisfactory up to $r_z=10$.
We use the structure functions \cite{mrs}.
The dependence of our results on this choice is negligible.

\begin{figure}[tbh]
\ \vspace{1cm}\\
\begin{minipage}[t]{7.8cm} {
\begin{center}
\hspace{-1.7cm}
\mbox{
\epsfysize=7.0cm
\epsffile[0 0 500 500]{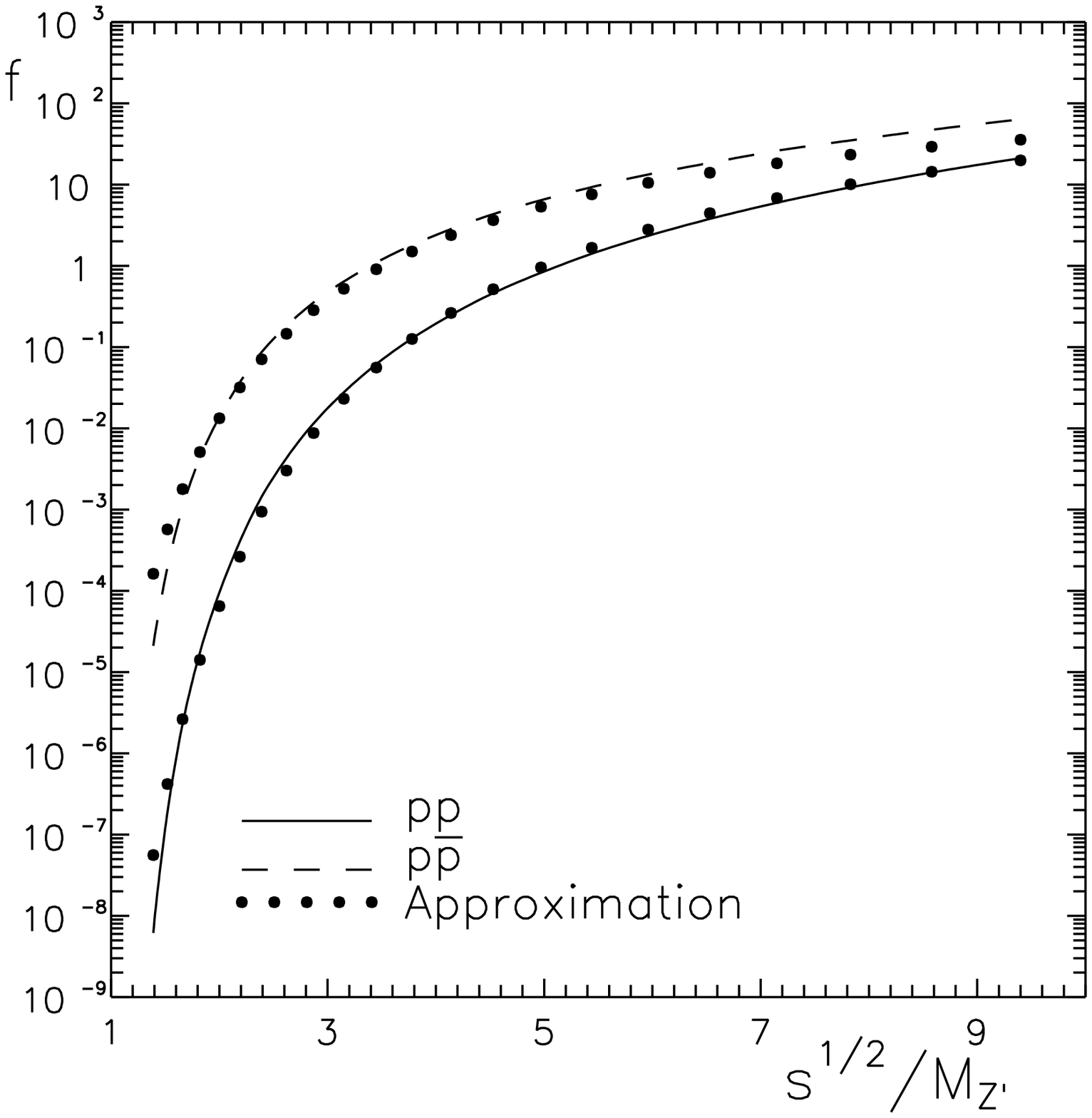}
}
\end{center}
\vspace*{-0.5cm}
\noindent
{\small\it {\bf Fig. 1:\ }
The function $f^u(\sqrt{s}/M_{Z'},25TeV^2)$ and the approximation
\req{fapprox2}.  
The curves of $f^u(\sqrt{s}/M_{Z'},Q^2)$ for $Q^2=600TeV^2$ or
$1TeV^2$ could not be distinguished from $f^u(\sqrt{s}/M_{Z'},25TeV^2)$.
}
}\end{minipage}
\hspace*{0.5cm}
\begin{minipage}[t]{7.8cm} {
\begin{center}
\hspace{-1.7cm}
\mbox{
\epsfysize=7.0cm
\epsffile[0 0 500 500]{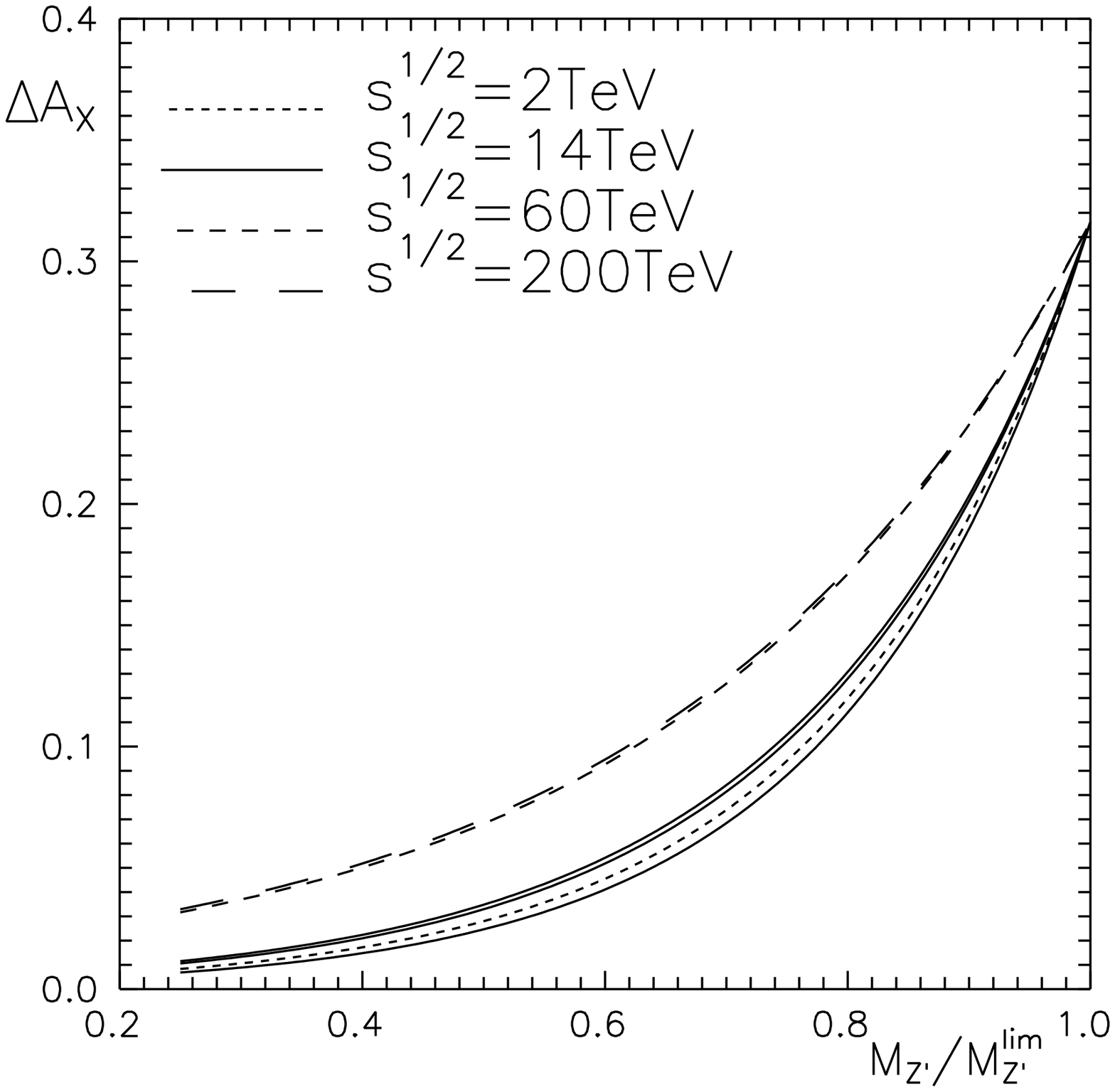}
}
\end{center}
\vspace*{-0.5cm}
\noindent
{\small\it {\bf Fig. 2:\ }
The estimate of $\Delta A_X$ as a function of
$M_{Z'}/M_{Z'}^{lim}$ as given in equation \req{fdef6} for $N_{Z'}=10$.
Shown are the predictions for the scenarios listed in table~1 for $Z'=\eta$.
For $\sqrt{s}=14TeV$, the dependence is shown for $Z'=\psi,\eta,SSM$
(from top to down). 
}
}\end{minipage}
\end{figure}

The expected number of $Z'$ events can now be written as
\ba
\label{fdef2}
N_{Z'}&=&L
\sigma_T\left(pp(p\bar p)\rightarrow (\nobodyfrac\gamma,Z,Z')X\rightarrow
\mu^+\mu^-X\right)
\approx\frac{L}{s}c_{Z'}C\exp\left\{-A\frac{M_{Z'}}{\sqrt{s}}\right\},\nll
&&\mbox{with\ \ } c_{Z'}=\frac{4\pi^2}{3}\frac{\Gamma_{Z'}}{M_{Z'}} 
Br(Z'\rightarrow \mu^+\mu^-)\left[
Br(Z'\rightarrow u\bar u) + \frac{1}{C_{ud}} Br(Z'\rightarrow d\bar d)\right]. 
\ea
All details of the $Z'$ model are collected in the constant $c_{Z'}$.
The approximate exponential dependence of $N_{Z'}$ on $M_{Z'}$ can be
recognized, for instance, from figures~1 to 5 in reference \cite{prd30}.
It also holds in associated $Z'$ production, $pp\rightarrow Z'W,
pp\rightarrow Z'Z$, as can be seen from figure~3 of reference \cite{prd46}. 

To predict $M_{Z'}^{lim}$, we have to invert equation \req{fdef2},
\bq
\label{fdef4}
M_{Z'}^{lim}\approx\frac{\sqrt{s}}{A}
\ln\left(\frac{L}{s}\frac{c_{Z'}C}{N_{Z'}}\right),
\eq
where now $N_{Z'}$ is the number of detected $Z'$ events demanded for a $Z'$
signal. 
Relation \req{fdef4} describes the scaling of $M_{Z'}^{lim}$ with the
c.m. energy and the integrated lminosity.
$M_{Z'}^{lim}$ depends on $L$ only logarithmically. 
Therefore, $M_{Z'}^{lim}$ depends only marginally on
detector efficiencies or event losses due to background suppression.
The model dependent constant $c_{Z'}$ enters \req{fdef4} only under the
logarithm leading to the weak model dependence of $Z'$ exclusion
limits in $pp$ and $p\bar p$ collisions.   
The physical origin of this effect is hidden in the properties of
the structure functions entering the definition \req{fdef} of $f(r_z)$.
Therefore, relation \req{fdef4} obtained for 
$\sigma_T\left(pp(p\bar p)\rightarrow (\nobodyfrac\gamma,Z,Z')X\rightarrow
\mu^+\mu^-X\right)$ is qualitatively true for other observables too.

Radiative corrections lead to deviations of $N_{Z'}$ from the Born
prediction, which can be taken into account by a multiplication with a
$K$ factor, $N_{Z'}\rightarrow KN_{Z'}$.
The effect on $M_{Z'}^{lim}$ is expected to be moderate because also $N_{Z'}$
enters this limit only under the logarithm.

The scaling \req{fdef4} can be
compared with the scaling law known for $e^+e^-$ collisions \cite{zpmi,zepp},
\bq
\label{epemlim}
M_{Z'}^{lim}\approx \left( sL\right)^{1/4}.
\eq
Not shown in \req{epemlim} is the direct proportionality to the square of the
coupling constants of the $Z'$ to SM fermions \cite{zpmi}, which is the
origin of the strong model dependence of $M_{Z'}^{lim}$.

For practical purposes, it is useful to write equation \req{fdef4} in
the form 
\bq
\label{ppzpscale}
\frac{M_{Z'}^{lim}(s,L)}{M_{Z'}^{lim}(s_0,L_{0})}\approx
\frac{\sqrt{s}}{\sqrt{s_0}}\left(1+\xi\ln\frac{s_0L}{sL_0}\right),
\hspace{1cm} \xi=\left[\ln\frac{L_0}{s_0}\frac{c_{Z'}C}{N_{Z'}}\right]^{-1},
\eq
where now all model dependence is hidden in the constant $\xi$.
Normalizing at one collider, equation \req{ppzpscale} predicts the limits for
colliders with different energy and luminosity.

All $Z'$ exclusion limits published in figure~1 of reference \cite{godfrey}
can be reproduced by equation \req{ppzpscale} 
with an accuracy of 10\% for 
$\xi=0.13\ (0.10)$ for $pp\ (p\bar p)$ collisions. 
The logarithmic dependence of $M_{Z'}^{lim}$ on $L$ can also be recognized
in figure~3 of reference \cite{rizzosnow}.
The reduction of $M_{Z'}^{lim}$ due to a decrease of the event rate by
a factor two is predicted by relation \req{ppzpscale} to be $9\%\ (7\%)$ for
$pp\ (p\bar p)$ collisions. 
These numbers, which do not discriminate between $Z'$ models and
colliders are in agreement with the last line of table~2
in reference \cite{rizzosnow}.

The analyses \cite{godfrey} and \cite{rizzosnow} report the search limits
for different future colliders for $N_{Z'}=10$, where it is assumed
that the $Z'$ does not decay to exotic fermions.
The analysis \cite{godfrey} is based on detected muon pairs only, while 
in \cite{rizzosnow} the electron pairs are included too.
We selected two scenarios from every paper giving $\sqrt{s}$ and $L$
in table~1. 
The numbers are produced from figure~1 of \cite{godfrey} and  taken 
from table~2 of \cite{rizzosnow}.
The $Z'$ models $\chi,\psi$ and $\eta$ in table~1 belong to the $E_6$
theory, $LR$ is a $Z'$ in the left-right symmetric model, while $SSM$ is the
$Z'$ in the sequential standard model. 
We see that the prediction \req{fdef4} agrees with the exact results
within 10\% in a wide range of $L$ and $s$.
At fixed $s$ and $L$, equation \req{fdef4} predicts for $pp$
collisions $M_{Z'}^{lim}/M_{SSM}^{lim}=.91(0.82,0.85,0.93)$ for 
$Z'=\chi(\psi,\eta,LR)$.
This prediction is in good agreement with the numbers quoted in table~1. 

%
\begin{table}[tbh]
\begin{center}
\begin{tabular}{|lrr|rrrrrr|}\hline\rule[-2ex]{0ex}{5ex} 
Analysis &$\frac{\sqrt{s}}{TeV}$ &$\frac{L}{fb^{-1}}$
&$\chi$&$\psi$ &$\eta$ &$LR$ &$SSM$ &estimate \req{fdef4}\\ 
\hline
\cite{godfrey} & 2($\ p\bar p$)&   10 & 1.05 & 1.05 & 1.08 & 1.10 & 1.15 & 
\rule[0ex]{0ex}{3ex} 1.06\\
\cite{godfrey} & 14($\ pp$)  &  100 & 4.46 & 4.15 & 4.30 & 4.54 & 4.80 & 4.47\\
\cite{rizzosnow} & 60($\ pp$)&  100 & 13.3 & 12.0 & 12.3 & 13.5 & 14.4 & 15.0\\
\cite{rizzosnow} & 200($\ pp$)& 1000 & 43.6 & 39.2 & 40.1 & 43.2 & 45.9 & 
\rule[-1ex]{0ex}{3ex} 49.3\\
\hline
$1000\cdot c_{Z'}(pp)$&   &  & 1.17 & 0.572 & 0.712 & 1.35 & 2.27 & $-$ \\
$1000\cdot c_{Z'}(p\bar p)$&&& 0.40& 0.437 & 0.556 & 0.77 & 1.41 &
\rule[-1ex]{0ex}{3ex} $-$
\\
\hline
\end{tabular}\medskip
\end{center}

{\small\it  {\bf Tab. 1:}
The lower bound on $Z'$ masses
$M_{Z'}^{lim}$ in TeV, which could be excluded by the different colliders.
The estimate \req{fdef4} for $M_{SSM}^{lim}$ is added in the last column.
The last two lines contain the values of $1000\cdot c_{Z'}$, for convenience.
The $Z'$ is assumed to decay into SM fermions only.
} \end{table}

If a $Z'$ signal is found in hadron collisions, one would like to measure 
some details of the $Z'$ model.
This can be done \cite{e6}, \cite{rizzosnow}-\cite{prd46} by
measurements of different asymmetries $A_X$. 
A reasonable model measurement requires enough events to assume that
they are gaussian distributed. 
The one-$\sigma$ statistical errors can then be estimated as 
$\Delta A_X\approx 1/\sqrt{N_{Z'}}$.
From equation \req{fdef2}, we deduce the following estimate of $\Delta A_X$,
\bq
\label{fdef5}
\Delta A_X \approx N_{Z'}^{-1/2}
\approx \sqrt{\frac{s}{L}\frac{1}{c_{Z'}C}}
\exp\left\{\frac{AM_{Z'}}{2\sqrt{s}}\right\}.
\eq
Relation \req{fdef5} relies on the approximation \req{fapprox2},
which becomes inaccurate for too large $\sqrt{s}/M_{Z'}$.
One should therefore be careful in interpolating to $\sqrt{s}/M_{Z'}>10$.
Compared to $M_{Z'}^{lim}$, the error of the asymmetry measurement,
$\Delta A_X$, is much more dependent on the $Z'$
model because the constant $c_{Z'}$ enters not under the logarithm.
The estimate \req{fdef5} can be confronted with the results quoted in
table~2 of reference \cite{prd46}, which are obtained 
for $\sqrt{s}=14TeV,\ L=100fb^{-1}$ and $M_{Z'}=1TeV$:
\bq
\begin{array}{cccccc}
\mbox{Model:} & \chi  & \psi & \eta & LR    & SSM\nll
\Delta A_X \mbox{\ from\ \req{fdef5}}:& 0.008 & 0.012& 0.011& 0.008 & 0.006\nll
\Delta A_{FB}^e \mbox{\ from\ \cite{prd46}}:& 0.007 & 0.016& 0.014& 0.006 & -
\end{array}\eq 
Having in mind the crude approximations, which lead to the estimate
\req{fdef5}, the agreement is good.

Combining equations \req{fdef4} and \req{fdef5}, we can predict
the precision of the measurement of $A_X$ for a given
$M_{Z'}<M_{Z'}^{lim}$  if we know only $M_{Z'}^{lim}$ for the same collider,
\bq
\label{fdef6}
\Delta A_X\approx 
\left(\sqrt{\frac{s}{L}\frac{1}{c_{Z'}C}}\right)^{1-M_{Z'}/M_{Z'}^{lim}}
\cdot\left(\sqrt{N_{Z'}}\right)^{-M_{Z'}/M_{Z'}^{lim}}.
\eq
The dependence \req{fdef6} is illustrated in figure~2.
It is similar for different colliders and $Z'$ models.

The approximations \req{fdef5} and \req{fdef6} relying on the
statistical errors only do not hold for
measurements of $M_{Z'}$ and $\Gamma_{Z'}$, where the systematic errors
become important. 
See reference \cite{prd45} for details.

To summarize, we have derived simple scaling laws for $Z'$ exclusion
limits and for statistical errors of $Z'$ asymmetry measurements in $pp$
and $p\bar p$ collisions.
Our estimates were confronted with existing exact results of $Z'$ analyses
and found to be in good agreement with them.
The estimates make the dependence of $Z'$ limits on the c.m. energy,
the luminosity and the $Z'$ model parameters transparent.
They are useful rules to extrapolate existing $Z'$ limits to other
colliders and $Z'$ models.

\centerline{\bf Acknowledgment}
I would like to thank T. Riemann for the careful reading of the manuscript.
%

\end{document}